# HIGHER SCHOOL OF ECONOMICS
NATIONAL RESEARCH UNIVERSITY

*Yetkin Çınar*

**RESEARCH AND TEACHING EFFICIENCIES OF TURKISH UNIVERSITIES WITH HETEROGENEITY CONSIDERATIONS: APPLICATION OF MULTI-ACTIVITY DEA AND DEA BY SEQUENTIAL EXCLUSION OF ALTERNATIVES METHODS**

The research and teaching efficiencies of 45 Turkish state universities are evaluated by using Multi-Activity Data Envelopment Analysis (MA-DEA) model developed by Beasley (1995). Universities are multi-purpose institutions, therefore they face multiple production functions simultaneously associated with research and teaching activities. MA-DEA allows assigning priorities and allocating shared resources to these activities. We also consider other possible reasons for the heterogeneity in the sample and use a novel approach proposed by Aleskerov & Petrushchenko (2013). This method can assess efficiency by constructing new frontiers taking into account heterogeneity. From the results generated by MA-DEA model "research efficient" and "teaching efficient" universities are identified. It is also shown that priorities given to the universities' activities have an influence on obtained rankings. It is concluded that various models which take into account heterogeneity are worth using for reliability of efficiency evaluations.

Keywords: Multi-Activity DEA, DEA by Sequential Elimination of Alternatives, Turkish Universities, joint-efficiency, heterogeneity

*Yetkin Çınar*, Asst. Prof. Dr., Faculty of Political Sciences, Ankara University, Ankara; Post-Doctoral Researcher, International Laboratory of Decision Choice and Analysis, National Research University, "Higher School of Economics", Moscow.

E-mail: ycinar@ankara.edu.tr; yetkincinar@gmail.com; telephone: +90 543 8430187; +7 926 4494685.

* The author thanks to Prof. Dr. Fuad Aleskerov & Vsevolod Petrushchenko for their valuable comments and contributions.

# 1. Introduction

In recent years, a considerable expansion has been observed in the higher education system of Turkey, many new universities and programs have been established. By the end of 2006 the number of public universities was 52 and today this number has been almost doubled and reached 103. During this process, measurement of efficiency in higher education has become one of the most important discussion topics.

Data Envelopment Analysis (DEA) proposed by Charnes et al. (1978) is a widely-used methodology for efficiency evaluation, since it provides an efficient frontier in terms of comparisons among decision making units (DMUs) in the existing sample. DEA yields a single dimensionless performance index for multi-product DMUs without a priori assumption of formal analytic production function or input-output prices. The production process of higher education institutions (HEIs) has a multi-input/multi-output nature as well. Besides, it is difficult to obtain input-output prices in non-profit organizations such as HEIs. Hence, DEA is a suitable tool for measuring efficiency in higher education.

The efficiencies of the universities in the United States, United Kingdom, Australia, Canada, Germany, Netherlands, Italy, China and Russian Federation have been evaluated by using DEA-based methodologies (see, for instance, Rhodes & Southwick (1986), Johnes (2006), Athanassopoulos & Shale (1997), Abbott & Doucouliagos (2003), Avkiran (2001), McMillan & Datta (1998),Warning (2004), Salerno (2006), Agasisti & Dal Bianco (2006), Johnes & Yu (2008), Abankina et al. (2012), Aleskerov & Petruschenko (2013)). For Turkey, there are also several studies which use DEA in efficiency evaluation of universities. Some recent studies are Köksal & Nalçacı (2006), Ustasüleyman (2007), Özden (2008), Çokgezen (2009), Çelik & Ecer (2009) and Ulucan (2011), among others.

Despite its popularity, DEA has some limits for use. For example, in standard DEA evaluation the sample is compared with a unique production frontier constructed by all efficient units in the sample. But this implicit assumption is not valid for some industries in which DMUs allocate resources to the production of different types of outputs. If there are DMUs which focus on different activities, then they might be efficient in some of the activities but not in others, and this causes the bias of standard DEA scores.

This issue becomes more important for HEIs due to some special characteristics of these institutions (Worthington, 2001). First, universities have more than one purpose, namely teaching, research and public/social services. Second, stakeholders of these institutions have different expectations, conflicting values and goals which suggest that while some of them are focusing on education, others try to improve their research performance. Hence, to obtain reliable results we should take into account different goals of a HEI's activities.

Beasley (1995) proposed a solution to this problem. He introduced a DEA-based model called "Multi-Activity DEA (MA-DEA)" and applied it for the evaluation of research and teaching efficiencies of physics and chemistry departments in UK universities. This method is suitable for the evaluation of institutions which face multiple production functions using common resources, since these resources can be objectively assigned to different activities by the resolution model itself. The model has been applied to the efficiency evaluation of multi-purpose, large-scaled, non-profit (state) institutions in such fields as education, health, police forces and bus services (see, e.g. Salerno (2006),Tsai & Mar Molinero (2002), Diez-Ticio & Mancebon (2002) and Yu (2007)).

In this study, the research and teaching efficiencies of state universities in Turkey are jointly taken into account and measured by using the MA-DEA model. For Turkey, all previous studies have used methodologies which take into account unique production processes.

On the other hand, heterogeneity problem may be considered not only in the sense of different goals. Being a crucial assumption of DEA, homogeneity implies that the DMUs undertake the same production processes and they operate within the same environment. One of the possible solutions to this issue is to construct different efficiency frontiers taking into account the heterogeneity in the sample. This idea is a basis for the novel methodology proposed by Aleskerov and Petrushchenko (2013).Their method is also applied to the Turkish dataset in this research.

Since different efficiency rankings are produced by the models mentioned above, the results then compared by using Kendall's τ (Kendall, 1938).

The text is organized as follows. Section 2 gives a brief explanation of the methodology. In Section 3 we define the basic parameters for evaluation, then the results of the empirical application are presented. Section 4 concludes.

# 2. Methodology

## 2.1. Data Envelopment Analysis (DEA)

DEA is a multi-factor productivity analysis model for measuring the relative efficiencies of a homogenous set of decision making units (DMUs). Formally, consider $S$ DMUs $\{1,..,S\}$ produce $n$ outputs $\{q_1,..q_i,.,q_n\}$ by using $m$ inputs $\{x_1,..x_j,.,x_m\}$. The optimal efficiency score for $k$-th DMU can be obtained by the following model

$$Max\, \theta_k = \frac{\sum_{i=1}^{n} u_i q_{ik}}{\sum_{j=1}^{m} v_j x_{jk}} = \frac{u_1 q_{1k} + u_2 q_{2k} + ... + u_n q_{nk}}{v_1 x_{1k} + v_2 x_{2k} + ... + v_m x_{mk}} \quad (1)$$

$$s.t. \quad \theta_s = \frac{\sum_{i=1}^{n} u_i q_{is}}{\sum_{j=1}^{m} v_j x_{js}} \leq 1, \quad s \in \{1,...,k,...,S\}, \quad (2)$$

$$\forall i,j \quad u_i, v_j \geq \varepsilon. \quad (3)$$

Here $u_i$'s and $v_j$'s are the weights prescribed to the outputs and inputs, respectively. The objective function in this model is the ratio of weighted sum of outputs to the weighted sum of inputs of the $k$-th university. The model finds the weights that maximize this ratio. Then the obtained value of $\theta_k$ represents the efficiency score of $k$-th DMU. This unit is called "efficient" if $\theta_k$ equals to 1, otherwise it is called "inefficient". The above problem is run $S$ times to compute the relative efficiency scores for each DMU in the sample.

The model dual to (1) – (3) was also introduced in Charnes et al. (1978). Both are valid under constant return to scale (CRS) assumption. Banker et al. (1984) modified the model by adding the variable return to scale (VRS) assumption.

## 2.2. Problem Formulation and The Multi-Activity DEA Model

In this study Multi-Activity DEA Model is adapted to the problem based on information about the available data (also see Section 3) for the Turkish Higher Education System. The production process which jointly takes into account teaching and research activities of Turkish universities is illustrated in *Fig. 1*. Here the MA-DEA model is briefly explained based on this structure. The more generalized mathematical formulation of the method can be found in Beasley (1995).

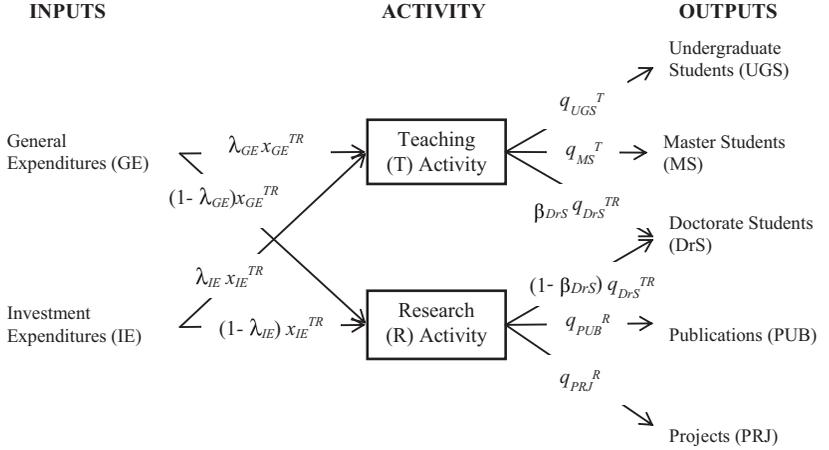

**Figure 1.** Multi-Activity (Research & Teaching) Production Process

According to *Fig.1*[1] the amounts of general and investment expenditures are distributed among teaching and research activities by the proportions of corresponding λ's. The numbers of undergraduate and master students are associated only with teaching activity. Similarly, the outputs solely associated with research activity are the number of publications and the total budget for (accepted) projects. According to the model, the number of doctorate students is distributed among teaching activity (with the weight $\beta_{DrO}$) and by research activity (with $1-\beta_{DrO}$). This assumption is made, since there is no available data about the research post-graduates and teaching post-graduates in Turkey.

According to this structure, multi-activity efficiency score for *k-th* university ($\theta_k$) can be obtained as a solution of the mathematical optimization problem

$$\text{Max } \theta_k = \alpha_k^T \cdot \theta_k^T + \alpha_k^R \cdot \theta_k^R \qquad (4)$$

---

[1] Here the upper indices T, R and TR show, respectively, that the variable is either associated with teaching, research or both activities. The lower indices stand for the abbreviations of the corresponding variables.

**s.t.   Teaching Efficiency** ($\theta_k^T$)

$$\frac{u_{UGS}\, q_{UGS,s}^T + u_{MS}\, q_{MS,s}^E + \beta_{DrS}\, u_{DrS}\, q_{DrS,s}^{TR}}{\lambda_{GE}\, v_{GE}\, x_{GE,s}^{TR} + \lambda_{IE}\, v_{IE}\, x_{IE,s}^{TR}} \leq 1,\ s = 1,...k...,S\,, \qquad (5)$$

**Research Efficiency** ($\theta_k^R$)

$$\frac{u_{PUB}\, q_{PUB,s}^R + u_{PRJ}\, q_{PRJ,s}^R + (1-\beta_{DrO})\, u_{DrO}\, q_{DrO,s}^{TR}}{(1-\lambda_{GE})\, v_{GE}\, x_{GE,s}^{TR} + (1-\lambda_{IE})\, v_{IE}\, x_{IE,s}^{TR}} \leq 1,\ s = 1,...k...,S\,, \qquad (6)$$

Priority Weights for Activities

$$\alpha_k^T + \alpha_k^R = 1, \qquad (7a)$$

$$\alpha_k^T = (\lambda_{GE}\, v_{GE}\, x_{GE,k}^{TR} + \lambda_{IE}\, v_{IE}\, x_{IE,k}^{TR}) / (v_{GE}\, x_{GE,k}^{TR} + v_{IE}\, x_{IE,k}^{TR})\,, \qquad (7b)$$

Limits for Variables

$$\forall u, v \geq \varepsilon\,;\ \forall s\ \theta_{s.}^T, \theta_{s.}^R, \theta_{s.} \geq 0\,;\ \lambda, \beta \geq \varepsilon\,, \qquad (8)$$

$$1 \geq \alpha_k^T \geq 0\,;\ 1 \geq \alpha_k^R \geq 0\,. \qquad (9)$$

Here $k$ and $s$ stand for DMUs, $\theta_k^T, \theta_k^R, \theta_k$ show teaching, research and total multi-activity efficiency scores of university $k$, respectively, and $v_{GE}, v_{IE}$, $u_{UGS}, u_{MS}, u_{PUB}, u_{PRJ}, u_{DrS}$ show the weight allocations to the corresponding inputs and outputs.

The MA-DEA model determines not only the optimal weights as in (1)-(3), but also the allocations of the shared variables which jointly maximizes the ratio of outputs to inputs used for teaching and research. Equations (5) and (6) break the equation (2) into separate teaching and research components. Then, as shown in the objective function (4), the model estimates *k-th* University's overall efficiency as a weighted average of the component efficiencies computed in (5) and (6). The weights ($\alpha$'s) in the objective function reflect the relative value the institution places on each activity. By this way, the model maximizes the efficiency of each production process separately and simultaneously. Since their sum is normalized to 1, the distribution of $\alpha$'s can also be interpreted as the priority (importance) given to each activity. In the above model, $\alpha$ is defined as a fraction of total weighted input source devoted to teaching for university $k$, following the definition given by Beasley (1995: 446). Here $\lambda's$ and $\beta's$ are the new variables to be estimated. To ob-tain a reliable result, these weights are restricted to be greater than some small positive value $\varepsilon$.

After Beasley (1995), the dual of the model is written and some mathematical properties are studied by Mar Molinero (1996) and Mar Molinero & Tsai (1997). A variable return to scale assumption is added to the model by Tsai & Mar Molinero (2002).

## *2.3. DEA by Sequential Exclusion of Alternatives*

Heterogeneity problem may be considered not only in the sense of different goals or activities. In this study, considering other possible reasons for the heterogeneity in the data set, the efficiencies are also computed by using a novel approach proposed by Aleskerov & Petrushchenko (2013). They offer a new method of efficiency estimation based on a sequential exclusion of alternatives in DEA calculations. The method allows assessing efficiency by constructing a new frontier taking into account heterogeneity of the evaluated sample. In their original paper, the authors focus on the heterogeneity caused by drastic differences in operating environment which means that many units can be located quite far from efficiency frontier. This type of heterogeneity causes that all inefficient units are benchmarked against some outstanding ones which are very rare in the whole sample. To overcome this issue they propose to move the efficiency frontier towards the barycenter of the units. Their approach can be illustrated via a graphical representation given in *Fig. 2* (Aleskerov & Petrushchenko, 2013, p. 9).

In *Fig. 2*, $F_1$ is the efficient unit according to standard DEA (CRS) model. As it can be seen, the DMUs $F_2$, $F_3$, $F_4$, $F_5$ and $F_6$ are benchmarked against $F_1$ which showed exceptional efficiency via standard DEA calculation. In other words, a vast majority of the units $F_3,…,F_6$ are located very far from the efficient frontier formed by $F_1$. According to the Aleskerov & Petrushchenko (2013)'s approach, these inefficient units should be evaluated less strictly than in the case of standard DEA model. For this aim, the barycenter (point B) of all units is calculated in the usual geometrical sense. The next step is to construct the new frontier via generating a unit G on the segment between the efficient DMU and the barycenter of the sample ($BF_1$ in *Fig.2*). Then $F_3, ..., F_6$ can be benchmarked against the unit G instead of $F_1$. Since $F_1$ and $F_2$ remain unevaluated, to compute their efficiency scores the same algorithm is repeated excluding the firms $F_3, ..., F_6$ from consideration.

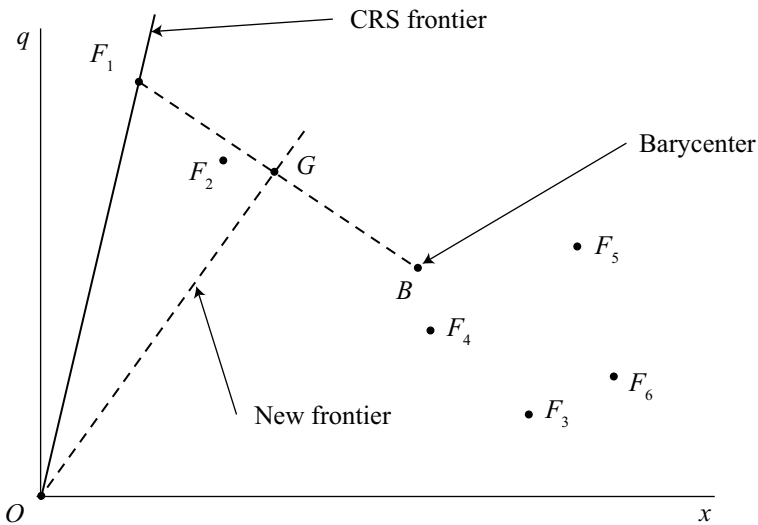

**Figure 2.** Graphic Interpretation of DEA by Sequential Exclusion of Alternatives Method

The location of the unit G should be coherent with some extent to heterogeneity. It means that the higher heterogeneity within the sample the nearer generator to the barycenter. An index μ is introduced for a heterogeneity degree of the sample which can be computed as the ratio of the mean value of $d$ to the maximum element in $d$, where the vector $d$ consists of Euclidian distances from the barycenter to all DMUs. It is noted that this value can also be computed in a different way, for instance, via expert evaluations.

Then the position of G is defined as

$G = \mu B + (1 - \mu) F_1$  (10)

The algorithm converges for any sample and the only DMU that remains efficient is the one which is efficient according to standard DEA model.

The algorithm of the method can be summarized as follows.

*Step 1:* Calculation of the standard DEA scores and the barycenter of all DMUs in data set.

*Step 2:* Move all efficient DMUs according to standard DEA towards the barycenter. The heterogeneity degree μ is needed at this stage of the model. Roughly speaking, the higher μ the higher heterogeneity.

*Step 3:* The calculation of new efficiency scores for all inefficient DMUs regarding the moved DMUs.

*Step 4:* All DMUs that obtain new efficiency scores less than 1 are excluded from the sample.

*Step 5:* The procedure goes from the top with the refreshed set of DMUs.

Further descriptions and more generalized formulation of the method can be found in Aleskerov & Petrushchenko (2013). The method has been applied to 29 Russian universities in the original paper. In our paper, we apply this methodology to Turkish data set and compare efficiency rankings produced by two studies.

## 3. Data, Application and Results

### 3.1. Data and Sample

Available input and output data for selected 45 state universities[2] for the year 2010 were obtained from The National Scientific and Technological Research Council (www.tubitak.gov.tr), The Council of Higher Education (www.yok.gov.tr), Student Selection and Placement Center (www.osym.gov.tr). Financial data is obtained from the Ministry of Finance General Directorate of Budget and Fiscal Control (www.bumko.gov.tr).

The descriptive statistics of the variables used in this research are given in Table 1.

*Table 1.* Descriptive Statistics of the Input / Output Variables

| Variables | Descriptive Statistics (No. of Observations = 45) | | | |
|---|---|---|---|---|
| **Inputs** | **Mean** | **Std. Error** | **Max.** | **Min.** |
| General Expenditures (GE) /.000 TL | 134,260 | 13,009 | 467,192 | 29,768 |
| Investment Expenditures (IE) /.000 TL | 28,659 | 1,543 | 68,264 | 13,000 |
| **Outputs** | **Mean** | **Std. Error** | **Max.** | **Min.** |
| No. of Undergraduate Students (UGS) | 27,799 | 2,103 | 70,405 | 1,806 |
| No. of Master Students (MS) | 2,460 | 340 | 9,429 | 476 |
| No. of Doctorate Students (DrS) | 890 | 168 | 4,832 | 102 |
| No. of Publications in Citation Indices (PUB) | 466 | 50 | 1570 | 30 |
| Total Budget for Accepted Projects (PRJ) /.000 TL | 3,103 | 625 | 25,079 | 218 |

---

[2] All established before the year 2006 and incorporate teaching and research activities.

From Table 1 it can be seen that the sample consists of relatively big universities and the smaller ones. However, all of them incorporate teaching and research activities by producing associated outputs in acceptable quantities.

Correlations among the variables are given in Table 2.

*Table 2.* Correlations between variables

| Variable | GE | IE | UGS | MS | DrS | PUB | PRJ |
|---|---|---|---|---|---|---|---|
| GE | – | | | | | | |
| IE | 0.45 | – | | | | | |
| UGS | 0.62 | 0.11 | – | | | | |
| MS | 0.81 | 0.23 | 0.64 | – | | | |
| DrS | 0.82 | 0.34 | 0.42 | 0.82 | – | | |
| PUB | 0.43 | 0.35 | 0.02 | 0.36 | 0.75 | – | |
| PRJ | 0.92 | 0.51 | 0.50 | 0.72 | 0.83 | 0.55 | – |

Table 2 shows that there are not very high correlations within inputs or within outputs, which can be interpreted as the selected variables are suitable and have individual meanings for DEA calculations.

### 3.2. Application

The application is conducted in the following steps.

**1)** *Calculation of Standard DEA Scores.* The classical DEA model given in (1)-(3) is applied to the data set and standard DEA (CRS) scores are obtained.

**2)** *Application of MA-DEA Model.* To jointly estimate teaching, research and total MA-DEA efficiency scores, the model (4)-(9) is applied by using two scenarios on the priority weights $\alpha's$ [3].

The choice of these priorities is important because different priority weights will lead different splits between resources allocated to the activities. This can also be seen as a matter of judgment similar to some multi-criteria decision models, as it is stated by Mar Molinero (1996, p. 1279). There are two alternative ways to choose weights. First, it is possible to decide for the values of $\alpha's$ externally, e.g. they might be coming from the strategic priorities of universities. When this information is not valid, they can be chosen on some other reasons, e.g. activities can be weighted equally. The other way is that the model determines these weights implicitly. It is one of the major benefits of this approach. The following two scenarios represent these two alternative ways of choosing priorities.

---

[3] The non-linear mathematical programming model given in (4)-(9) is written and solved by LINGO®. The program consists of 101 variables, 10 of them being non-linear, and it consists of 93 constraints (91 of them being non-linear).

*Scenario 1.* In this scenario $\alpha$ parameters are equal, i.e. $\alpha_k^T = \alpha_k^R = 0.5$ which means that both activities are equally prioritized by all universities. Note that in this scenario constraint (7b) is not used. It is considered as a base scenario, aiming to compare the results with the standard DEA's results objectively. However, in order not to have degenerated results, the following definitions are made: $\forall u, v \geq \varepsilon$ ($\varepsilon = 0.00001$) and $0.99 \geq \{\lambda, \beta\} \geq 0.01$.

*Scenario 2.* In this scenario, the model is expected to calculate the $\alpha$'s itself, implying the optimal importance weight attachments of each university. That is, $\alpha$'s are considered as variables defined in constraint (7b). To get closer to the real cases, the assumption $0.90 \geq \{\lambda, \beta\} \geq 0.30$ is added. In this way, the possibility of zero allocations to any activity of shared variables or priorities is eliminated.

**3)** *Application of DEA by Sequential Eliminations of Alternatives Method*

The model is applied to the data set of Turkish universities by assigning $\mu = 0.2$, $\mu = 0.5$ and $\mu = 0.8$ as in the original application of the model to the Russian universities' data set. In other words, in both applications the same heterogeneity levels are used.

### *3.3. Results*

The efficiency scores obtained by standard DEA calculations and via the application of MA-DEA model are given in Appendix 1 and summarized in Table 3.

Results first show that MA-DEA model generates different efficiency scores from the standard DEA calculation.

MA-DEA results are significantly lower than the classical ones, both individually and in terms of mean values (see Table 3 and Appendix 1). While the standard model identifies 8 efficient universities, there are 2 "research efficient" (U17, U37) and 4 "teaching efficient" (U15, U32, U33, U39) universities, which are only efficient in one activity but not in the other. Two universities (U21 and U37) remain efficient in both activities; they were also efficient in standard DEA calculations. However, all universities which are efficient in only one activity by the MA-DEA models were efficient in standard

DEA model as well. These results can be explained by the fact that achievement of maximum efficiency by MA-DEA model requires to be efficient in both of the component activities simultaneously.

*Table 3.* Summary of Results obtained from CRS-DEA and MA-DEA Models

| Summary of Results | DEA (CRS) | MA-DEA | | | | | | |
|---|---|---|---|---|---|---|---|---|
| | | Scenario 1 $\alpha_s^E = \alpha_s^A = 0.5$ | | | Scenario 2 $\alpha_s^E$ and $\alpha_s^A$ variable | | | |
| | | MA-DEA (Total) | Teaching Activity | Research Activity | $\alpha_s^E$ | MA-DEA (Total) | Teaching Activity | Research Activity |
| Mean Efficiency Score | 0.86 | 0.66 | 0.65 | 0.67 | 0.55 | 0.72 | 0.64 | 0.67 |
| Std. Dev. of Efficiency Scores | 0.11 | 0.13 | 0.21 | 0.17 | 0.32 | 0.14 | 0.21 | 0.18 |
| No. of Efficient Universities | 8 | 2 | 6 | 4 | - | 2 | 6 | 4 |

There are 17 universities which have higher teaching efficiency scores than their research efficiencies, and 26 universities have higher research efficiency scores (see Appendix 1). Research efficiency is higher than teaching efficiency with respect to mean values as well. Thus one can conclude that in Turkish Higher Education System research efficiency is higher than teaching efficiency.

When priorities are freely selected by universities (Scenario 2), it is observed that by assigning higher priorities to the activities in which the universities are more efficient, they can reach higher total efficiencies. Different priorities given to activities not only result in different efficiency rankings, but also make the strategic behavior of the universities to be observable on the weight allocations. From the detailed outputs of the model, it can be observed how universities attach weights to their inputs and outputs, and how they allocate the resources to different activities according to their assigned priorities. It is important when the relation between performance analysis and strategic planning is considered.

By assuming different frontiers for different activities, the results obtained by MA-DEA model give a more clear perspective. In other words, since in the standard model only one production function is considered, and substitutions between activities are quite possible, generated results can be biased.

Finally, Table 4 shows the results obtained from the application of 'DEA by Sequential Exclusion of Alternatives' model to the Turkish data set and presents a comparison of all rankings obtained by all models in terms of Kendall's distances (Kendall, 1938)[4].

*Table 4.* Comparison of Efficiency Rankings

| Kendall's Distances | | Turkish Universities | Russian Universities |
|---|---|---|---|
| | | **DEA (CRS)** | **DEA (CRS)** |
| **DEA by Sequential Exclusion of Alternatives** | **μ = 0.2** | 0.062 | 0.020 |
| | **μ = 0.5** | 0.141 | 0.049 |
| | **μ = 0.8** | 0.221 | 0.116 |
| **MADEA - Scenario 1** | | 0.168 | – |
| **MADEA - Scenario 2** | | 0.128 | – |

Although the Turkish and Russian datasets and configurations of the applied models are not fully comparable, the results in Table 4 show that

– the heterogeneity degree in Turkish data set is higher than in the Russian Universities' data set (for all values of $\mu's$),

– the assertion of Aleskerov & Petrushchenko (2013)'s model "the higher $\mu$ the higher heterogeneity" is verified for Turkish data set as well,

– there is significant difference between MA-DEA rankings and the original ones.

More generally, these results show that both methods capture the heterogeneity in some degree and in some sense. Therefore, a limitation of classical DEA calculation is shown.

## 4. Conclusion

In this study, measuring the efficiency in Turkish Higher Education System, we have considered two approaches which define different production functions and efficient frontiers compared with the standard DEA model. While the Multi-Activity DEA model assumes separate production functions for different activities of DMUs, the 'DEA by Sequential Exclusion of Alter-natives' model evaluates efficiency by constructing new frontiers taking into account heterogeneity at different levels. It can be stated that since all data sets are in some sense heterogeneous, the models which define different frontiers are worth considering for reliability of efficiency evaluations.

---

[4] The value of Kendall's distance lie between 0 and 1, where 0 means that two orderings are the same and 1 means that the two rankings are inverse.

The results show that important differences exist between the universities with respect to their research and teaching activities' efficiency. The multi-activity DEA model is capable of identifying sources of inefficiency by means of different activities. It is also shown that priorities given to the universities' activities have an influence on obtained rankings. Hence, the estimated results can provide one to propose strategies for improving operational performance of the inefficient universities by means of assigned priorities and source allocations. Thereby MA-DEA model potentially yields greater managerial insights into organizational improvements than classical DEA models.

The results obtained from the application of 'DEA by Sequential Exclusion of Alternatives' model to the Turkish universities' data set has shown that the heterogeneity degree of the sample is high (and higher than the Russian data set). This can be caused by differences between the regions which universities are located, by differences in scales or due to the other reasons. Therefore an extension of this research can be made by controlling the effects of these factors to the efficiency.

Finally, evaluating multi-activity efficiencies of universities focusing on social sciences or natural sciences might be another extension of this study.

# Appendix 1. DEA and MA-DEA Efficiency Results *

| University | DEA | Multi Activity DEA (MA-DEA) | | | | | | |
| --- | --- | --- | --- | --- | --- | --- | --- | --- |
| | | *Scenario-1* $\alpha_s^T = \alpha_s^R = 0.5$ | | | *Scenario-2* $\alpha_s^T$ and $\alpha_s^R$ (variable) | | | |
| | Effi-ciency | MA-DEA Total Eff. | MA-DEA Teach-ing Eff. | MA-DEA Research Eff. | $\alpha_s^T$ | MA-DEA Total Eff. | MA-DEA Teach-ing Eff. | MA-DEA Research Eff. |
| U1 | 0.77 | 0.56 | 0.56 | 0.56 | 0.86 | 0.56 | 0.56 | 0.56 |
| U2 | 0.87 | 0.61 | 0.59 | 0.63 | 0.32 | 0.62 | 0.59 | 0.63 |
| U3 | 0.97 | 0.71 | 0.81 | 0.61 | 0.90 | 0.77 | 0.81 | 0.44 |
| U4 | 0.74 | 0.51 | 0.38 | 0.64 | 0.30 | 0.56 | 0.37 | 0.64 |
| U5 | 0.85 | 0.79 | 0.74 | 0.84 | 0.30 | 0.80 | 0.70 | 0.84 |
| U6 | 0.88 | 0.82 | 0.75 | 0.88 | 0.36 | 0.84 | 0.75 | 0.88 |
| U7 | 0.98 | 0.71 | 0.95 | 0.48 | 0.90 | 0.89 | 0.93 | 0.46 |
| U8 | 0.87 | 0.64 | 0.50 | 0.78 | 0.30 | 0.69 | 0.48 | 0.78 |
| U9 | 0.83 | 0.62 | 0.63 | 0.61 | 0.86 | 0.63 | 0.63 | 0.61 |
| U10 | 0.69 | 0.54 | 0.65 | 0.43 | 0.90 | 0.62 | 0.65 | 0.33 |
| U11 | 0.95 | 0.68 | 0.65 | 0.70 | 0.32 | 0.68 | 0.64 | 0.70 |
| U12 | 0.74 | 0.65 | 0.65 | 0.65 | 0.53 | 0.65 | 0.65 | 0.65 |
| U13 | 0.69 | 0.45 | 0.30 | 0.61 | 0.30 | 0.51 | 0.28 | 0.61 |
| U14 | 0.98 | 0.77 | 0.83 | 0.70 | 0.90 | 0.82 | 0.83 | 0.70 |
| U15 | **1.00** | 0.75 | **1.00** | 0.50 | 0.90 | 0.95 | **1.00** | 0.50 |
| U16 | 0.88 | 0.72 | 0.57 | 0.88 | 0.30 | 0.75 | 0.43 | 0.88 |
| U17 | **1.00** | 0.84 | 0.68 | **1.00** | 0.30 | 0.88 | 0.61 | **1.00** |
| U18 | 0.73 | 0.51 | 0.38 | 0.64 | 0.30 | 0.56 | 0.38 | 0.64 |
| U19 | 0.85 | 0.61 | 0.48 | 0.73 | 0.30 | 0.65 | 0.47 | 0.73 |
| U20 | 0.63 | 0.59 | 0.63 | 0.20 | 0.90 | 0.59 | 0.63 | 0.20 |
| U21 | **1.00** | **1.00** | **1.00** | **1.00** | 0.30 | **1.00** | **1.00** | **1.00** |
| U22 | 0.81 | 0.56 | 0.41 | 0.72 | 0.30 | 0.62 | 0.40 | 0.72 |
| U23 | 0.79 | 0.56 | 0.54 | 0.59 | 0.32 | 0.57 | 0.54 | 0.59 |
| U24 | 0.79 | 0.58 | 0.37 | 0.79 | 0.30 | 0.66 | 0.36 | 0.79 |
| U25 | 0.63 | 0.47 | 0.38 | 0.56 | 0.30 | 0.50 | 0.37 | 0.56 |
| U26 | 0.97 | 0.71 | 0.67 | 0.76 | 0.30 | 0.72 | 0.63 | 0.76 |
| U27 | 0.84 | 0.79 | 0.73 | 0.84 | 0.34 | 0.79 | 0.71 | 0.84 |
| U28 | **1.00** | 0.68 | 0.36 | **1.00** | 0.30 | 0.81 | 0.36 | **1.00** |
| U29 | 0.85 | 0.57 | 0.47 | 0.66 | 0.30 | 0.60 | 0.47 | 0.66 |
| U30 | 0.94 | 0.78 | 0.82 | 0.74 | 0.85 | 0.80 | 0.81 | 0.73 |
| U31 | 0.92 | 0.62 | 0.55 | 0.69 | 0.30 | 0.64 | 0.53 | 0.69 |
| U32 | **1.00** | 0.77 | **1.00** | 0.54 | 0.90 | 0.95 | **1.00** | 0.50 |
| U33 | **1.00** | 0.87 | **1.00** | 0.74 | 0.90 | 0.97 | **1.00** | 0.72 |
| U34 | 0.79 | 0.59 | 0.61 | 0.56 | 0.90 | 0.61 | 0.61 | 0.54 |
| U35 | 0.79 | 0.57 | 0.70 | 0.45 | 0.90 | 0.67 | 0.70 | 0.43 |
| U36 | 0.92 | 0.67 | 0.43 | 0.92 | 0.30 | 0.76 | 0.40 | 0.92 |
| U37 | **1.00** | **1.00** | **1.00** | **1.00** | 0.30 | **1.00** | **1.00** | **1.00** |
| U38 | 0.83 | 0.60 | 0.61 | 0.59 | 0.86 | 0.61 | 0.61 | 0.59 |
| U39 | **1.00** | 0.71 | **1.00** | 0.42 | 0.90 | 0.94 | **1.00** | 0.42 |
| U40 | 0.99 | 0.80 | 0.93 | 0.66 | 0.90 | 0.89 | 0.93 | 0.54 |
| U41 | 0.94 | 0.71 | 0.70 | 0.71 | 0.39 | 0.70 | 0.69 | 0.71 |
| U42 | 0.68 | 0.52 | 0.57 | 0.46 | 0.90 | 0.55 | 0.57 | 0.39 |
| U43 | 0.86 | 0.76 | 0.86 | 0.66 | 0.90 | 0.83 | 0.86 | 0.56 |
| U44 | 0.82 | 0.65 | 0.63 | 0.67 | 0.40 | 0.65 | 0.62 | 0.67 |
| U45 | 0.68 | 0.46 | 0.33 | 0.60 | 0.30 | 0.51 | 0.31 | 0.60 |
| **Mean** | **0.86** | **0.66** | **0.65** | **0.67** | **0.55** | **0.72** | **0.64** | **0.67** |
| **Standard Deviation** | **0.11** | **0.13** | **0.21** | **0.17** | **0.32** | **0.14** | **0.21** | **0.18** |

(*) The scores of the efficient universities are emphasized by using bold characters.

**Чинар, Е.** Эффективность исследований и преподавания в турецких университетах с учетом неоднородности: использование многозадачного оболочечного анализа данных и оболочечного анализа данных с последовательным исключением альтернатив [Текст] : препринт WP7/2013/04 / Е. Чинар ; Нац. исслед. ун-т «Высшая школа экономики». – М.: Изд. дом Высшей школы экономики, 2013. – 24 с. – 15 экз. (на англ. яз.). – (Серия WP7 «Математические методы анализа решений в экономике, бизнесе и политике»).

Оценена эффективность исследований и преподавания в 45 турецких университетах с использованием многозадачного оболочечного анализа данных (МЗ-ОАД), разработанного в Beasley (1995). Университеты – это многоцелевые организации, описываемые многокритериальной целевой функцией, связанной с исследованиями и обучением. МЗ-ОАД позволяет учесть приоритеты и распределить общие ресурсы между составляющими этой функции. Мы рассматриваем и другие возможные причины возникновения неоднородности в выборке и используем новый подход, предложенный в Aleskerov & Petrushchenko (2013). Этот метод позволяет оценить эффективность путем построения серии новых границ, учитывая степень неоднородности выборки. Из результатов, полученных по методу МЗ-ОАД, выявлены эффективные в смысле исследований и в смысле преподавания университеты. Показано, что приоритеты, приписываемые различным видам деятельности, влияют на полученные ранжирования. Показано, что использование различных моделей, учитывающих неоднородность, важно для надежности оценок эффективности.

Ключевые слова: многозадачный метод оболочечного анализа данных, оболочечный анализ данных с последовательным исключением альтернатив, турецкие университеты, общая эффективность, неоднородность

*Препринт WP7/2013/04*
*Серия WP7*
Математические методы анализа решений
в экономике, бизнесе и политике

Еткин Чинар

**Эффективность исследований и преподавания
в турецких университетах с учетом неоднородности:
использование многозадачного оболочечного
анализа данных и оболочечного анализа данных
с последовательным исключением альтернатив**

(*на английском языке*)